\documentclass{elsart}
\journal{Physics Letters A}
\usepackage{graphicx,bm,amsmath,amssymb}
\newcommand{\rl}{\rangle\!\langle}
\DeclareMathOperator{\tr}{Tr}

\begin{document}
\begin{frontmatter}

\title{``Which path'' decoherence in quantum dot experiments}

\author{K. Roszak},
\ead{katarzyna.roszak@pwr.wroc.pl}
\author{P. Machnikowski}
\address{Institute of Physics, 
Wroc{\l}aw University of Technology, 50-370 Wroc{\l}aw, Poland}

\begin{abstract}
We analyze and interpret 
recent optical experiments with semiconductor quantum dots.
We derive
a quantitative relation between the amount of information transferred
into the environment
and the optical polarization
that may be observed in a spectroscopy experiment.
\end{abstract}
 
\end{frontmatter}

\maketitle

\section{Introduction}

The fundamental difficulty in sustaining the coherence of a
quantum system is its interaction with the surrounding world. In the
course of joint evolution of the system and its environment, this
interaction establishes phase correlations between the former and a
macroscopic number of degrees of freedom of the latter. For most states, 
such correlations
perturb and eventually erase the internal phase information of the
quantum system turning a non-classical superposition state into a mixture
of a small subset of classicaly allowed states. This effect is known as \textit{decoherence} or
\textit{dephasing} and seems to be one of the most fundamental aspects
of the quantum theory \cite{zurek03,joos03}. It is also of importance for
certain practical tasks, like quantum computing
\cite{nielsen00}, where maintaining system coherence over
many control operations is indispensable.

Once the quantum correlations (entanglement) 
between the system and its environment have been
created, a measurement on the environment
may, in principle, yield information on the system. 
Thus, one may say that a certain amount of information, allowing one
to determine the system state, has been 
transferred to the environment. Since both information
transfer and decoherence result from the same underlying physical
process of building up correlations, a certain
relation between the amount of information transfer
and the phase coherence retained in the system should be expected.
The most celebrated textbook example of such a relation is the vanishing of
interference fringes in a double-slit experiment with single particles
whenever one tries to establish through which slit the particle passes. 
Because of this historical relation to the
interference of two spatial paths, the knowledge of the system state
is customarily referred to as 
``which path'' (or \textit{welcher Weg}) information.

The decoherence effect due to the transfer of information to the
environment appears always
when a quantum system is coupled to the surrounding world. In this
paper we revisit one specific example of this effect: the decay of
coherent optical polarization in a quantum dot due to the dephasing of
confined carrier states, which we will interprete as a result of
information leakage from the carrier subsystem to its environment
(phonon modes). After introducing the system and defining its model
(Sec.~\ref{sec:model}) we will derive the evolution of the confined
carriers coupled to phonons using an algebraic technique based
on Weyl operators
(Sec.~\ref{sec:dephas}). Next, we will 
discuss a quantitative measure of the distinguishability of
quantum states due to the information contained in the environment
(Sec.~\ref{sec:dist}). Finally, in Sec.~\ref{sec:complem}, we will use
the facts introduced in the preceding sections to show that the pure
dephasing effect observed as the decay of the coherent response from
an optically excited semiconductor quantum dot (or an ensemble
thereof) \cite{borri01} may be interpreted as a result of
\textit{which path} dephasing. We derive a quantitative
complementarity relation between the observed coherent optical
response and the amount of information transferred to the environment.

%The paper is organized as follows. 
%In the next Section we formulate the model of the carrier-phonon system.
%In Sec.~\ref{sec:dephas} we describe the system evolution and the
%resulting damping of optical polarization. In Sec.~\ref{sec:dist}
%we define a quantitative measure of state distinguishability.
%Section \ref{sec:complem} 
%contains the derivation of the complementarity relation
%which is the central part of this paper. 
%Sec.~\ref{sec:concl} concludes the paper.

\section{The system and the model}
\label{sec:model}

In this paper, we consider the simplest version of a time-resolved optical
experiment performed on a single quantum dot (for practical reasons,
actual experiments are often performed on ensambles of QDs using
nonlinear techniques \cite{borri01}). A very short laser pulse 
prepares the system in a certain superposition (dependent on the pulse
phase and intensity) of the ground state
(no exciton, denoted $|0\rangle$) and the single-exciton state
(denoted $|1\rangle$). By \textit{very short} we mean a pulse 
much shorter than the time scales of phonon
dynamics, so that the preparation of the initial state may be
considered instantaneous. This corresponds to the actual experimental
situation with pulse durations of order of $100$ fs \cite{borri01}. On the
other hand, the pulse is long enough to assure a relatively narrow
spectrum and prevent the population of higher confined levels. Let us
restrict (for simplicity) to the equal superposition state
\begin{equation}\label{equal}
|\psi_0\rangle=\frac{|0\rangle +|1\rangle}{\sqrt{2}}.
\end{equation}
In such a state, the inter-band component of the
electric dipole moment has a non-vanishing average value
oscillating at an optical frequency (hence referred to as
\textit{optical polarization}) which leads to the emission
of coherent electromagnetic radiation with
an amplitude proportional to the oscillating dipole moment. In an
unperturbed system (e.g., in an atom), the radiation
would be emitted over time of
the order of the lifetime of the superposition state, i.e., until the
system relaxes to the ground state due to radiative energy loss. 

In a semiconductor structure an additional effect, related to
carrier-phonon coupling, appears on a time scale much shorter than the
lifetime of the state.  
Due to the interactions between confined carriers and lattice ions, 
the ground state of the lattice in the
presence of a charge distribution is different than in its
absence. As a result, after the creation of a confined exciton 
the lattice 
relaxes to a new equilibrium, which is accompanied by the emission of
phonon wave packets \cite{jacak03b,vagov02a} that form a trace in the
macroscopic crystal distinguishing the exciton state from an empty
dot. As we will discuss below, this information broadcast via emitted
phonons leads to a decay of the coherence of the superposition state
although the average
occupations of the system states remain unaffected (hence the process
is referred to as \textit{pure dephasing}).
Since coherent dipole radiation requires well-defined phase relations 
between the components
of a quantum superposition, the amplitude of this radiation, measured in
the experiment, gives access to the coherence properties of the
quantum state of confined carriers itself. The dephasing of the qantum
superposition is therefore directly translated into the decay of coherent
optical radiation from the system.

For the carrier-phonon system in a semiconductor quantum dot, 
a microscopic model of the dephasing effect exists (involving the
interaction with acoustic phonons), which reproduces
experimental data very well \cite{vagov03,vagov04} and may serve
as a reliable starting point to describe the evolution of the
combined system of confined carriers and lattice modes. 

The Hamiltonian of the system is
\begin{equation}\label{ham}
H=\epsilon|1\rl 1|+H_{\mathrm{ph}}
+|1\rl 1|\sum_{\bm{k}}
(f_{\bm{k}}^{*}b_{\bm{k}}+f_{\bm{k}}b_{\bm{k}}^{\dag}),
\end{equation}
where the first term describes the energy of the confined exciton
($\epsilon$ is the energy difference between the states without phonon
corrections),
$H_{\mathrm{ph}}
=\sum_{\bm{k}}\hbar\omega_{\bm{k}}b_{\bm{k}}^{\dag}b_{\bm{k}}$
is the Hamiltonian of the phonon subsystem and the 
third term describes the interaction. 
Carrier-phonon interaction constants in (\ref{ham})
are given by
\begin{equation}\label{cpl}
    f_{\bm{k}}
    =( \sigma_{\mathrm{e}} -\sigma_{\mathrm{h}} )
\sqrt{\frac{\hbar k}{2\varrho V_{\mathrm{N}} c}}
\int_{-\infty}^{\infty} d^3\bm{r}\psi^*(\bm{r})
e^{-i\bm{k}\cdot\mathrm{r}}\psi(\bm{r}),
\end{equation}
and describe the deformation potential coupling between the
carriers and the lattice modes, which is the dominating mechanism
under the assumed optical driving conditions \cite{krummheuer02}.
Here $\varrho$ is the crystal density, $V_{\mathrm{N}}$ is the
normalization volume of the phonon system, $\omega_{\bm{k}}=ck$ is the
frequency of the phonon mode with the wave vector $\bm{k}$ 
($c$ is the speed of longitudinal sound), and 
$b_{\bm{k}}^{\dag}$, $b_{\bm{k}}$ are 
phonon creation and annihilation operators. The exciton wave function
is modelled as a product of two identical single-particle wave
functions $\psi(\bm{r}_{\mathrm{e}})$ and $\psi(\bm{r}_{\mathrm{h}})$,
corresponding to the electron and hole, respectively.

In our calculations we use typical parameters for a self-assembled
InAs/GaAs structure: single particle wave functions 
$\psi(\bm{r})$ modelled by
Gaussians with $4$ nm width in the $xy$ plane and $1$ nm along $z$, 
the deformation
potential difference $\sigma_{\mathrm{e}}-\sigma_{\mathrm{h}}=9.5$ eV,
crystal density $\varrho=5300$ kg/m$^{3}$, and the speed of
longitudinal sound $c=5150$ m/s.

\section{Phonon-induced pure dephasing 
of optical polarization}
\label{sec:dephas}

In this Section we describe the dephasing of a confined exciton state
within the exactly solvable model of interaction with the
environment presented above. 
After describing the time evolution of the system we
find the degree of coherence
remaining in the system as manifested by the amplitude of the
experimentally measurable coherent dipole radiation.
In this way we reproduce the recent theoretical
description \cite{krummheuer02} using a simple algebraic method which
provides the complete density matrix of the carrier subsystem,
necessary for the discussion to be presented in Sec. \ref{sec:complem}.

The carrier-phonon interaction term in Eq.~(\ref{ham}) 
is linear in phonon operators
and describes a shift of the lattice equilibrium induced by the
presence of a charge distribution in the dot. The stationary state of
the system corresponds to the exciton and the surrounding coherent 
cloud of phonons representing the lattice distortion to the new
equilibrium. The transformation that creates the coherent cloud is
the shift 
$wb_{\bm{k}}w^{\dagger}=b_{\bm{k}}-f_{\bm{k}}/(\hbar\omega_{\bm{k}})$,
generated by the Weyl operator (see Appendix)
\begin{equation}
\label{w}
w= 
\exp\left[\sum_{\bm{k}} 
\left( \frac{f_{\bm{k}}}{\hbar\omega_{\bm{k}}}b_{\bm{k}}^{\dag}
-\frac{f_{\bm{k}}^{*}}{\hbar\omega_{\bm{k}}} b_{\bm{k}} \right)\right].
\end{equation}
A straightforward calculation shows that the Hamiltonian (\ref{ham})
is diagonalized by the unitary transformation 
$W=|0\rl 0|\otimes \mathbb{I}+|1\rl 1|\otimes w$,
where $\mathbb{I}$ is the identity operator and the tensor product
refers to the carrier subsystem (first component) and its phonon
environmnent (second component). As a result one gets
\begin{displaymath}
\widetilde{H} = W H W^{\dagger}
= E|1\rl 1|+H_{\mathrm{ph}},
\end{displaymath}
where 
$E=\epsilon-\sum_{\bm{k}}|f_{\bm{k}}|^{2}/(\hbar\omega_{\bm{k}})$.

We assume that 
at the beginning ($t=0$) the state of the whole system is
$\rho_{0}=(|\psi_{0}\rangle\!\langle \psi_{0} |)\otimes \rho_{\mathrm{E}}$,
where $\rho_{\mathrm{E}}$ is the density matrix of the phonon
subsystem (environment) at thermal equilibrium and
$|\psi_{0}\rangle$ is the equal superposition state (\ref{equal})
prepared by a properly chosen ultrashort pulse.

The evolution operator $U(t)=e^{-iHt/\hbar}$ may
be written as
\begin{displaymath}
U(t)=W^{\dagger}WU(t)W^{\dagger}W
= W^{\dagger}\widetilde{U}(t)W 
= W^{\dagger} \widetilde{U}(t) W\widetilde{U}^{\dagger} (t)
\widetilde{U}(t) =W^{\dagger}  W(t)\widetilde{U}(t),
\end{displaymath}
where $ \widetilde{U}(t) =e^{-i\widetilde{H}t/\hbar}$ and 
$ W(t)=\widetilde{U}(t) W\widetilde{U}^{\dagger} (t)$.
Since $\widetilde{U}(t)$ is
 diagonal the explicit form of $W(t)$ may easily be found
and one gets
\begin{equation}
\label{op_phon}
U(t)=\left[ |0\rl 0|\otimes \mathbb{I}
+|1\rl 1|\otimes w^{\dag}w(t) \right]\widetilde{U}(t),
\end{equation}
where $w(t)=e^{-iH_{\mathrm{ph}}t/\hbar}we^{iH_{\mathrm{ph}}t/\hbar}$.

Using the evolution operator in the form (\ref{op_phon}) the system
state at a time $t$ may be written as
\begin{equation}
\label{mac_czas}
\rho(t)=\frac{1}{2}
\left(
\begin{array}{cc}
\rho_{\mathrm{E}}&
e^{iEt/\hbar} \rho_{\mathrm{E}}  w^{\dagger}(t) w\\
e^{-iEt/\hbar}w^{\dagger} w(t)\rho_{\mathrm{E}} &
w^{\dagger} w(t)\rho_{\mathrm{E}}  w^{\dagger}(t) w
\end{array}
\right),
\end{equation}
where we used the tensor product notation in which an
operator $A$ is expanded as $A=\sum_{m,n}|m\rl n|\otimes A_{mn}$ with
a set of operators $A_{mn}$ acting on the second subsystem, and
written in the matrix form with respect to the first subsystem.
The density matrix for the carrier subsystem is obtained by tracing
out the phonon degrees of freedom, i.e., 
$\rho_{\mathrm{S}}=\tr_{\mathrm{E}}\rho$. Hence,
\begin{equation}
\label{dm-T}
\rho_{\mathrm{S}}(t)
=\frac{1}{2}\left(
\begin{array}{cc}
1& e^{iEt/\hbar}\langle  w^{\dagger}(t) w \rangle \\
e^{-iEt/\hbar}\langle w^{\dagger} w(t) \rangle&
1
\end{array}
\right),
\end{equation}
where the average may be calculated by first
using the multiplication rule for Weyl operators
[Eq.~(\ref{multi})] to combine $w^{\dag}$ and $w(t)$ and then 
Eq.~(\ref{srednia_k}). The result is
\begin{displaymath}
\langle  w^{\dagger}(t) w \rangle=
\exp\left\{-\sum_{\bm{k}} \left|\frac{f_{\bm{k}}}
{\hbar\omega_{\bm{k}}}\right|^{2}
\left[ i\sin\omega_{\bm{k}}  t 
+(1-\cos\omega_{\bm{k}}  t)(2n_{\bm{k}} +1)\right]\right\},
\end{displaymath}
where $n_{\bm{k}}$ are bosonic equilibrium occupation numbers.

\begin{figure}[tb] 
\begin{center} 
\unitlength 1mm
\begin{picture}(48,26)(0,3)
\put(0,0){\resizebox{53mm}{!}{\includegraphics{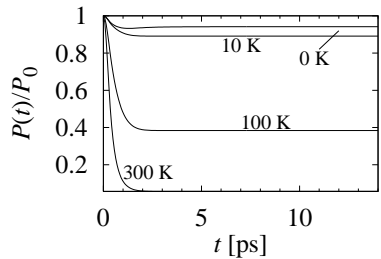}}}
\end{picture} 
\end{center} 
\caption{\label{fig:res_0}
Decay of the coherent radiation from a confined exciton at various
temperatures, as shown.} 
\end{figure}

The emitted coherent dipole
radiation is 
proportional to the non-diagonal element of the density matrix 
$\rho_s (t)$ and its amplitude is
\begin{equation}
\label{P}
P(t) = P_{0}|\langle  w^{\dagger}(t) w \rangle|.
\end{equation}
In Fig.~\ref{fig:res_0} we show the
normalized polarization amplitude $P(t)/P_{0}$
(first derived in Ref. \cite{krummheuer02}). The interaction
with the macroscopic crystal environment leads to a reduction of
coherent radiation due to pure dephasing of the exciton state,
reflected by the reduced value of the non-diagonal element of the
density matrix $\rho_{\mathrm{S}}$. 
At $t=0$ one has $\langle  w^{\dagger}(t) w \rangle=1$, while at
large values of $t$, $\cos\omega_{\bm{k}}t$
oscillates very quickly as a function of $\bm{k}$ 
and averages to 0 (see also Ref. \cite{krummheuer02}). 
Thus, for long times, the polarization
amplitude tends to a temperature-dependent finite value 
\begin{displaymath}
P(t)\to P_{0}\exp\left[-\sum_{\bm{k}}
\left|\frac{f_{\bm{k}}}{\hbar\omega_{\bm{k}}}\right|^{2}
(2n_{\bm{k}}+1)\right]<P_{0}.
\end{displaymath}
This partial decay of coherence
is a characteristic feature of short-time
dephasing for carrier-phonon couplings encountered in real systems
\cite{borri01}.

On the other hand, it is clear from the diagonal elements in 
Eq.~(\ref{mac_czas}) that the two carrier states $|0\rangle$ and
$|1\rangle$ are accompanied by different states of the phonon
environment so that correlations between the two subsystems are present.
These correlations may be related to the phonon wave packets that are
emitted into the bulk of the crystal after ultrafast optical
excitation \cite{vagov02a,jacak03b}.
These wave packets traveling away from the QD region carry the information
about the system state into the environment.
We will see that
dephasing may be quantitatively related to the amount of these
correlations, i.e., to the information
on the exciton state extracted by its phonon environment.
First, however, we need a quantitative measure for this information.

\section{Distinguishability of quantum states}
\label{sec:dist}

For completeness, in this Section we review the definition of 
a measure for the quantity of \textit{which path} information on
a quantum state of the system contained
in its environment \cite{englert96}.  

Qualitatively, the definition is as follows. Let us consider a
two-level system S with the Hilbert space spanned by the states
$|0\rangle,|1\rangle$, interacting with its environment E. A which
path measurement of the system S, e.g., a measurement of the
observable $|0\rl 0|$, determines the system state. Given the state of
the total system S$+$E, how well can we predict the outcome of this
measurement based on a previous measurement on the environment? The
measure of distinguishability is based on the probability $p$ of a
correct guess for the best possible choice of the measurement on the
environment. If the systems are completely uncorrelated then any
measurement on the environment is of no help and the probability of
correct prediction remains equal to 1/2. If the systems are in a
maximally entangled state like
$|0\rangle|E_{0}\rangle+|1\rangle|E_{1}\rangle$ with $\langle
E_{0}|E_{1}\rangle=0$ then the result of the measurement of the
environment observable $|E_{0}\rl E_{0}|$ completely determines the
state of the system S, so that $p=1$. The distinguishability measure
is defined as
\begin{equation}\label{disting-measure}
\mathcal{D}=2\left(p-\frac{1}{2}\right)
\end{equation}
and changes correspondingly from 0 to 1.

In general, for a given state of the total system $\rho$ and a result
$e$ of the 
environment measurement, the optimal prediction procedure
\cite{wooters79} is the usual statistical inference rule:
one calculates the probabilities 
\begin{equation}\label{pir}
p(i,e) = \tr \left[ (|i\rl i|\otimes |e\rl e|)\rho  \right]
\end{equation}
of finding the system S in the state $|i\rangle$ simultaneously with
the environment in the state $|e\rangle$ associated with the
measurement outcome $e$. Then, given the outcome $e$,
 one opts for the state $i$ for which this
probability is larger. The prediction is correct if the total system
is indeed in this more probable state, whichever it is. 
Allowing for all possible outcomes $e$, the total likelihood of
guessing right is equal to the probability that the system is in any
of the ``more probable states'', i.e,
\begin{displaymath}
p = \sum_{e}\max\left[ p(0,e),p(1,e) \right]
= \frac{1}{2}\sum_{e}\left[ p(0,e)+p(1,e) \right] 
+\frac{1}{2}\sum_{e}\left| p(0,e)-p(1,e) \right|.
\end{displaymath}
The first term contains the sum of probabilities for any state and is
therefore equal to 1/2.
Writing the full density matrix (for equal probabilities of the two
system states) 
in the form $\rho=(1/2)\sum_{ij}|i\rl j|\otimes \rho_{ij}$,
where $\rho_{00}$ and $\rho_{11}$ (but neither 
$\rho_{01}$ nor $\rho_{10}$) are density matrices,
and using Eq.~(\ref{pir}) we can write
\begin{displaymath}
p=\frac{1}{2}+\frac{1}{4}\sum_{e}
\left|\langle e| \rho_{00}-\rho_{11}|e\rangle  \right|.
\end{displaymath}

This formula gives the probability of correct prediction for a fixed
measurement on the environment, i.e. for a specific basis of environment 
eigenstates $\{|e\rangle\}$. In order to measure the amount of information
inherently contained in the environment, independently of the possibly
poor choice of the measurement basis, one maximizes this quantity with respect
to all possible observables that can be measured, i.e. to all
possible complete eigensystems $\{|e\rangle\}$. To find the
optimal (upper bound) value for $p$, we split the
hermitian operator 
$\Delta\rho=\rho_{00}-\rho_{11}$ into a positive and
negative part. Let $\Delta\rho_{l}$ and $|l\rangle$ be the
eigenvalues and the corresponding eigenstates of $\Delta\rho$. 
Let $L_{+},L_{-}$ be the sets of quantum numbers $l$ for which 
$\Delta\rho_{l}$ is nonnegative and negative, respectively. Then 
$\Delta\rho=\Delta\rho^{(+)}-\Delta\rho^{(-)}$, with the
positive operators 
$\Delta\rho^{(\pm)}=\sum_{l\in L_{\pm}} \Delta\rho_{l}|l\rl l|$.
Since these operators are positive, one has 
\begin{displaymath}
\left|\langle e|\Delta\rho^{(+)}|e\rangle
-\langle e|\Delta\rho^{(-)}|e\rangle \right|\le
\langle e|\Delta\rho^{(+)}|e\rangle
+\langle e|\Delta\rho^{(-)}|e\rangle
= \langle e\left| |\Delta\rho| \right|e\rangle,
\end{displaymath}
where $|\Delta\rho|$ is the modulus of the operator
$\Delta\rho$, defined as 
\begin{displaymath}
|\Delta\rho|
=\sum_{l}|\Delta\rho_{l}||l\rl l|=\Delta\rho^{(+)}+\Delta\rho^{(-)}. 
\end{displaymath}
Hence, 
\begin{displaymath}
\frac{1}{4}\sum_{e}\left|\langle e| \Delta\rho|e\rangle
\right|\le
\frac{1}{4}\tr |\Delta\rho|
\equiv \frac{1}{2}D(\rho_{00},\rho_{11}),
\end{displaymath}
where $D(\rho,\rho')$ is known as the trace distance \cite{nielsen00} 
between the density matrices $\rho$ and $\rho'$.
On the other hand, the equality is clearly attained if the eigensystem
$\{|e\rangle\}$ coincides with the eigenstates of $\Delta\rho$, i.e.,
when the observable measured on the environment is $\Delta\rho$
itself.
Thus, using the definition (\ref{disting-measure}), we finally get
\begin{equation}\label{dist-trdist}
\mathcal{D}=D(\rho_{00},\rho_{11}).
\end{equation}

\section{The complementarity relation}
\label{sec:complem}

In this Section we derive a quantitative complementarity
relation between the degree of coherence in the quantum dot system, as
manifesteed by the 
amplitude of coherent radiation described in  Sec.~\ref{sec:dephas},
and the amount of \textit{which path} information
transferred to the environment, as defined in Sec.~\ref{sec:dist}. 
This relation is analogous to the
visibility-distinguishability relation in the double-slit setup
\cite{englert96,jaeger95}. 

For the carrier-phonon state of Eq.~(\ref{mac_czas}) the
distinguishability of carrier states due to the correlations with the
phonon environment is, from Eq.~(\ref{dist-trdist}),
\begin{displaymath}
\mathcal{D}(t)=D(\rho_E,w^{\dagger} w(t)
\rho_{\mathrm{E}}  w^{\dagger}(t) w).
\end{displaymath}
One has, in general \cite{nielsen00}, 
$D^{2}(\rho,\rho')\le 1-F^{2}(\rho,\rho')$, where 
$F(\rho,\rho')\!=\!\tr\sqrt{\rho^{1/2}\rho'\rho^{1/2}}$ is the fidelity
measure of the distance between the states $\rho$ and $\rho'$. Hence
\begin{equation}
\label{jeden}
\mathcal{D}^{2}(t)\leq
1-\left[ \tr \sqrt{\rho_E^{1/2}
w^{\dagger} w(t)\rho_E w^{\dagger}(t) w
\rho_E^{1/2}} \right]^2
=1-\left[ \tr \left|\rho_E^{1/2}
w^{\dagger} w(t) \rho_E^{1/2}\right| \right]^2.
\end{equation}
Since $\tr|A|\geq |\tr A|$ we may write
\begin{equation}
\label{dwa}
\tr |\rho_E^{1/2}w^{\dagger} w(t)\rho_E^{1/2}| \geq 
|\tr (\rho_E w^{\dagger} w(t))| 
= |\langle w^{\dagger} w(t)\rangle|=P(t)/P_{0},
\end{equation}
where we used Eq.~(\ref{P}).
Combining Eqs.~(\ref{jeden}) and (\ref{dwa}) leads to the relation
\begin{equation}
\label{vd}
\left[ \frac{P(t)}{P_{0}} \right] ^2+\mathcal{D}^2(t)\leq 1,
\end{equation}
which shows that the relative decay of polarization is related to the
\textit{which path} information transfer to the environment. Note that
for pure states $\rho_{\mathrm{E}}$,
equality holds in Eqs.~(\ref{jeden}) and (\ref{dwa}),
so that  the relation (\ref{vd}) also turns into an equality.

\section{Conclusion}
\label{sec:concl}

In this paper, we have interpreted the dephasing effects observed in 
recent optical experiments on semiconductor quantum dots in terms of
``which path'' information transfer to the environment. 
Coherence properties of carriers
confined in a QD may be tested by detecting coherent
radiation emitted from the system as a result of optical polarization
induced by optical excitation. Due to the coupling between the system and
its phonon environment, in the course of quantum evolution lattice
modes get excited, which leads to the dephasing of the carrier state
manifested by a decrease of the amplitude of this radiation. 
We have derived an inequality between the
remaining amplitude of emitted radiation and the amount of
information carried to the bulk of the macroscopic crystal 
by lattice excitations.
From this point of view the dephasing of optical polarization 
is of the same nature as the well-known disappearance of interference
fringes in a double-slit (Young) experiment due to the \textit{which
path} knowledge. 

The complementarity relation (\ref{vd}) is analogous to similar
relations between the distinguishability of paths and the visibility
of interference fringes in the double-slit setup
\cite{englert96,jaeger95}. In fact, 
an interference experiment analogous to the double-slit experiment
has been performed using optically excited states of semiconductor
quantum dots \cite{bonadeo98}, where the two states
(analogous to \textit{paths}) are the absence or
presence of a single, optically created exciton, 
confined by the binding potential of the QD. 
It may be shown that the visibility of interference fringes in these
time-domain interference experiments is governed by the same
non-diagonal element of the reduced density matrix as the coherent
polarization in our discussion.

Our discussion confirms that incompatibility of the quantum behavior
and which path information is a general feature of the quantum 
world. We have shown that a manifestation of this general property 
may be identified in an experimental situation that is extremely
different from the historical concept of two spatial paths in a 
double-slit experiment.

\section*{Acknowledgments}
Supported by the Polish MNI (PB~2~P03B~08525).

\appendix
\section*{Appendix}
\section{Rules for Weyl operators}
%\label{sec:Weyl}

Applying the Weyl operator (\ref{w}) to the phonon annihilation 
operator $ b_{\bm{k}} $ 
and expanding the result we get\cite{mahan00}
\begin{displaymath}
w b_{\bm{k}}w^{\dag}=b_{\bm{k}}+ [S,b_{\bm{k}}] +
\frac{1}{2} [S,[S,b_{\bm{k}}]] +\dots ,
\end{displaymath}
where $S\!=\!\sum_{\bm{k}}( g_{\bm{k}} b^{\dagger}_{\bm{k}}
-g^*_{\bm{k}}b_{\bm{k}})$, $g_{\bm{k}}\!=\!f_{\bm{k}}/(\hbar\omega_{\bm{k}})$.
Since $[S,b_{\bm{k}}]\!=\!-g_{\bm{k}}$ and $[S,[S,b_{\bm{k}}]]\!=\!0$
(as all the higher commutators),
we obtain 
the shift effect of the	Weyl operator,
$w b_{\bm{k}} w^{\dagger}=b_{\bm{k}} -g_{\bm{k}}$.

Multiplying Weyl operators is done using the Baker-Hausdorff formula 
for separating operator exponents
\cite{mahan00}
\begin{equation}\label{CBH}
e^{A+B}=e^Ae^Be^{-[A,B]/2},
\end{equation}
where $A$ and $B$ are some operators and 
$[A,[A,B]]=[B,[A,B]]=0$.
Some simple algebra using Eq.~(\ref{CBH}) 
shows that the product of two Weyl operators, 
\begin{equation}
\label{i12}
w_i=\exp\left[\sum_{\bm{k}}\left(g_{\bm{k}}^{(i)}b_{\bm{k}}^{\dagger}
-g_{\bm{k}}^{(i)*}b_{\bm{k}}\right)\right], 
\end{equation}
$i=1,2$,
 is equal to 
\begin{equation}
\label{multi}
w_1 w_2= w_3 \exp\left[-\frac{1}{2}\sum_{\bm{k}}\left(g^{(1)*}_{\bm{k}} 
g^{(2)}_{\bm{k}} -g^{(1)}_{\bm{k}}g^{(2)*}_{\bm{k}}\right)\right],
\end{equation}
where $w_3$ is given by (\ref{i12}) with 
$g^{(3)}_{\bm{k}}=g^{(1)}_{\bm{k}}+g^{(2)}_{\bm{k}}$.

Now it is possible to calculate the equilibrium average of a Weyl 
operator (following Ref. \cite{mahan00}), 
$\langle w \rangle =\tr (\rho_{\mathrm{E}} w)$,
where $\rho_{\mathrm{E}}$ is the density matrix of the phonon 
reservoir at thermal equilibrium.
Since $\rho_{\mathrm{E}}=e^{-\beta H_{\mathrm{ph}}} /\tr 
e^{-\beta H_{\mathrm{ph}}}$, $\beta=1/(k_{\mathrm{B}}T)$,
and $H_{\mathrm{ph}}= 
\sum_{\bm{k}}\hbar\omega_{\bm{k}} b_{\bm{k}}^{\dagger} b_{\bm{k}} $, 
we may write
\begin{displaymath}
\langle w \rangle=
\frac{\tr \left( e^{-\beta \sum_{\bm{k}}\hbar\omega_{\bm{k}} 
b _{\bm{k}} ^{\dagger} b_{\bm{k}}}
e^{\sum_{\bm{k}}(g_{\bm{k}}b_{\bm{k}}^{\dagger}-g_{\bm{k}}^*b_{\bm{k}})}\right)}
{\tr (e^{-\beta \sum_{\bm{k}}\hbar\omega_{\bm{k}} 
b _{\bm{k}} ^{\dagger} b_{\bm{k}}})},
\end{displaymath}
and separate the thermal equilibrium into
\begin{equation}
\label{rozbicie}
\langle w \rangle=\prod_{\bm{k}} \langle w \rangle_{\bm{k}} ,
\end{equation}
where
\begin{equation}
\label{srednia_k}
\langle w \rangle_{\bm{k}}=
\frac{\tr \left( e^{-\beta \hbar\omega_{\bm{k}} 
b _{\bm{k}} ^{\dagger} b_{\bm{k}}}
e^{(g_{\bm{k}}b_{\bm{k}}^{\dagger}-g_{\bm{k}}^*b_{\bm{k}})}\right)}
{ \tr (e^{-\beta \hbar\omega_{\bm{k}} b _{\bm{k}} ^{\dagger} b_{\bm{k}}}) }.
\end{equation}
Using Eq.~(\ref{CBH}) and keeping in mind that
$\tr (e^{-\beta \hbar\omega_{\bm{k}} b _{\bm{k}} ^{\dagger} 
b_{\bm{k}}})=(1-e^{-\beta\hbar\omega _{\bm{k}} }) ^{-1},$
we can transform Eq.~(\ref{srednia_k}) into
\begin{equation}
\label{wuka}
\langle w \rangle_{\bm{k}}
=e^{\frac{1}{2}|g_{\bm{k}} |^2}
(1-e^{-\beta\hbar\omega _{\bm{k}} })
\tr (e^{-\beta \hbar\omega_{\bm{k}} b _{\bm{k}} ^{\dagger} b_{\bm{k}}}
e^{g_{\bm{k}}b_{\bm{k}}^{\dagger}}e^{-g_{\bm{k}}^*b_{\bm{k}}}) .
\end{equation}

For the trace one has
\begin{equation}
\label{slad}
\tr ( e^{-\beta \hbar\omega_{\bm{k}} b _{\bm{k}} ^{\dagger} b_{\bm{k}}}
e^{g _{\bm{k}} b_{\bm{k}}^{\dagger}}e^{-g_{\bm{k}}^*b_{\bm{k}}}) =
\sum_{ m =0}^{\infty}e^{-\beta \hbar\omega_{\bm{k}}m} 
\langle m |
e^{g _{\bm{k}} b_{\bm{k}}^{\dagger}} e^{-g_{\bm{k}}^*b_{\bm{k}}} |m\rangle ,
\end{equation}
where $|m \rangle$ denotes a state with $m$ excitations.
The exponents are expanded in a power series,
\begin{equation}
\label{rozw}
e^{-g_{\bm{k}}^* b_{\bm{k}} } |m\rangle=\sum_{l=0}^{\infty}
\frac{(-g^*_{\bm{k}})^{l}}{l!}b_{\bm{k}}|m\rangle .
\end{equation}
Since 
$b_{\bm{k}}^l|m\rangle= [m!/(m-l)!]^{1/2} |m-l\rangle$
for $l\leq m$ and $b_{\bm{k}}^l|m\rangle=0$ for $l>m$,
Eq.~(\ref{rozw}) transforms into
\begin{equation}
e^{-g_{\bm{k}}^* b_{\bm{k}} } |m\rangle = \sum_{l=0}^{m}
\frac{(-g^*_{\bm{k}})^{l}}{l!}\left[\frac{m!}{(m-l)!}\right]^{1/2} |m-l\rangle .
\end{equation}
Treating the other exponent in the same manner we end up with
\begin{equation}
\langle m|e^{g_{\bm{k}} b_{\bm{k}}^{\dagger} } e^{-g_{\bm{k}}^* b_{\bm{k}} } |m\rangle=
\sum_{l=0}^{m}
\frac{(-|g_{\bm{k}}|^2)^{l}}{(l!)^2}\frac{m!}{(m-l)!} ,
\end{equation}
that is, a Laguerre polynomial of order $m$, $L_m(|g_{\bm{k}}|^2)$.  
Using the fact that
\[
\sum_{m=0}^{\infty}L_m( |g_{\bm{k}}|^2 )z^m=
(1-z)^{-1}e^{ |g_{\bm{k}}|^2 \frac{z}{z-1}} ,
\]
with a specific $z= e^{-\beta\hbar\omega_{\bm{k}}} $ 
[so that $n_{\bm{k}}=z/(1-z)]$,
and inserting the result into Eq.~(\ref{wuka}), and then into 
Eq.~(\ref{rozbicie}),
we get the thermal average of the Weyl operator in the form
\begin {equation}
\label{weyl_sr}
\langle w \rangle=e^{-\frac{1}{2} \sum_{\bm{k}}|g_{\bm{k}} |^2(2n_{\bm{k}}+1)}.
\end{equation}

\bibliographystyle{prsty}
\bibliography{abbr,quantum}

\end{document}